\documentclass[aps,twocolumn]{revtex4-1}

\usepackage{graphics}
\usepackage{epsfig}
\usepackage{amsmath}

\def\ov#1{\overline{#1}}

\def\vb#1{\mbox{\boldmath$#1$}}
\def\pd#1#2{\frac{\partial #1}{\partial #2}}

\def\wh#1{\widehat{#1}}
\def\bdot{\,\vb{\cdot}\,}
\def\btimes{\,\vb{\times}\,}

\def\bhat{\wh{{\sf b}}}
\def\cal#1{\mathcal{#1}}

\def\eq#1{\eqref{eq:#1}}

\newcommand{\bc}{\begin{center}}
\newcommand{\ec}{\end{center}}
\newcommand{\bt}{\begin{tabbing}}
\newcommand{\et}{\end{tabbing}} 
\newcommand{\be}{\begin{eqnarray*}}
\newcommand{\ee}{\end{eqnarray*}}
\newcommand{\bs}{\begin{slide}}
\newcommand{\es}{\end{slide}}

\begin{document}

\title{Classical {\it zitterbewegung} in reduced plasma dynamics}

\author{Alain J.~Brizard}
\affiliation{Department of Chemistry and Physics, Saint Michael's College, Colchester, VT 05439, USA}

\begin{abstract}
The process of dynamical reduction of the Vlasov-Maxwell equations leads to the introduction of classical {\it zitterbewegung} effects in reduced plasma dynamics. These effects manifest themselves in the form of an asymmetric canonical energy-momentum tensor involving the decoupling of the reduced kinetic momentum ${\bf p}$ from the reduced velocity ${\bf u}$ (i.e., ${\bf u}\btimes{\bf p} \neq 0$) as well as reduced polarization and magnetization effects. The reduced intrinsic torque generated by the antisymmetric part of the canonical energy-momentum tensor, which is calculated from the reduced ponderomotive potential, acts as the source for the intrinsic (spin) angular momentum.
\end{abstract}

\begin{flushright}
October 27, 2010
\end{flushright}


\maketitle

In 1930, Schroedinger \cite{Schroedinger} provided a simple interpretation for the electron's spin angular momentum based on Dirac's  theory of the electron \cite{Dirac}. According to Schroedinger's {\it zitterbewegung} model \cite{Huang,Barut_Bracken,Barut_Zanghi}, the intrinsic spin angular momentum $S^{\mu\nu} \equiv \hbar\,\sigma^{\mu\nu}/2$ of the electron can be interpreted as the result of the rapid trembling motion of the electron. The main features of the Lagrangian formulation of {\it zitterbewegung} can be summarized by: (I) the decoupling of a particle's velocity ${\bf v}$ from its momentum ${\bf p}$ (i.e., ${\bf v}\btimes{\bf p} \neq 0)$; (II) the existence of polarization and magnetization effects associated with the decoupling between fast and slow space-time scales; and (III) an asymmetric canonical energy-momentum tensor derived from the Lagrangian density. In the Schroedinger {\it zitterbewegung} model, these features are explicitly related to the electron's spin. 

{\it Zitterbewegung} effects have recently been investigated in many different physical systems ranging from graphene \cite{GN_Nature,Castro_RMP} and photonic materials \cite{Dreisow_PRL,Zawadzki} to ultracold atoms \cite{Vaishnav_PRL} and cold-ion traps \cite{Lamata_PRL,Gerritsma_Nature}. The purpose of the present Letter is to show that the reduced Vlasov-Maxwell equations, which are obtained by dynamical reduction, exhibit these same
{\it zitterbewegung} features derived from the reduced ponderomotive Hamiltonian.

We begin with a brief review of the Schroedinger {\it zitterbewegung} model of the electron, which is governed by the Dirac-Maxwell equations 
\cite{RQM} 
\begin{equation}
\left. \begin{array}{rcl}
\gamma^{\mu}\,D_{\mu}\psi & = & mc^{2}\,\psi \\
\ov{D}_{\mu}\ov{\psi}\,\gamma^{\mu} & = & -\,mc^{2}\,\ov{\psi} \\
\partial_{\mu}F^{\mu\nu} & = & (4\pi/c)\,J^{\nu}
\end{array} \right\},
\label{eq:DM_eqs}
\end{equation}
where $J^{\nu} \equiv (c\varrho, {\bf J}) = ec\,\langle\gamma^{\nu}\rangle$ denotes the electron four-current (expressed in terms of the expectation value
$\langle\gamma^{\nu}\rangle \equiv \langle\psi|\gamma^{\nu}|\psi\rangle$ of the Dirac matrix $\gamma^{\nu}$), $D_{\mu} \equiv i\hbar\,\partial_{\mu} - (e/c)\,A_{\mu}$ and $\ov{D}_{\mu} \equiv i\hbar\,\partial_{\mu} + (e/c)\,A_{\mu}$ are the kinetic-momentum operators acting on $\psi$ and its adjoint 
$\ov{\psi}$, respectively, and $F_{\mu\nu} \equiv \partial_{\mu}A_{\nu} - \partial_{\nu}A_{\mu}$ denotes the electromagnetic field. We now show that the Lagrangian formulation of the Dirac-Maxwell equations \eq{DM_eqs} exhibit the three {\it zitterbewegung} features (I)-(III). 

First, the electron is described in terms of the coordinates ${\bf r}$, its velocity ${\bf v} \equiv d{\bf r}/dt$, and its kinetic momentum ${\bf p}$. While the electron's velocity has the eigenvalues $\pm\,c$, its expectation value $\langle{\bf v}\rangle$ (expressed in terms of the group velocity of its wavepacket) is less than $c$ in magnitude \cite{GN_Nature}. The motion of the electron is therefore decomposed into the slow (average) motion of its center of mass and the rapid trembling motion of its {\it zitterbewegung} position $\vb{\rho}_{\rm e}$. This rapid motion yields an expectation value for the magnetic dipole-moment of the electron \cite{Huang,Barut_mu,Barut_Zanghi}
\begin{equation}
\vb{\mu}_{\rm e} \;\equiv\; \frac{e}{2c}\;\left\langle\vb{\rho}_{\rm e}\btimes\frac{d\vb{\rho}_{\rm e}}{dt}\right\rangle,
\label{eq:mu_electron}
\end{equation}
whose magnitude $|\vb{\mu}_{\rm e}| \equiv e\hbar/(2mc)$ is the Bohr magneton.

Second, we discuss the electron's polarization and magnetization and introduce the Gordon decomposition of the electron four-current \cite{RQM}
\begin{equation}
\left. \begin{array}{rcl}
\varrho & \equiv & \varrho_{\rm c} - \nabla\bdot{\bf P} \\
 &  & \\
{\bf J} & \equiv & {\bf J}_{\rm c} + \partial{\bf P}/\partial t + c\,\nabla\btimes{\bf M}
\end{array} \right\},
\label{eq:rho_J_elec}
\end{equation}
where $J_{\rm c}^{\nu} \equiv (e/2m)\,[\langle D^{\nu}\rangle - \langle\ov{D}^{\nu}\rangle]$ denotes the ({\it spinless}) conduction four-current and the antisymmetric magnetization tensor 
\begin{equation}
M^{\mu\nu} \;\equiv\; \frac{e\hbar}{2m}\;\langle\sigma^{\mu\nu}\rangle 
\label{eq:M_Dirac}
\end{equation}
is expressed in terms of the spin matrix $\sigma^{\mu\nu} \equiv (i/2)\,(\gamma^{\mu}\gamma^{\nu} - \gamma^{\nu}\gamma^{\mu})$, whose components are the polarization $M^{i0} \equiv c\,P^{i}$ and the magnetization $M^{ij} \equiv \epsilon^{ijk}\,c\,M_{k}$.

Third, the canonical energy-momentum tensor for the Dirac-Maxwell equations
\begin{equation}
T^{\mu\nu} \;=\; g^{\mu\nu} \left( \frac{{\sf F}:{\sf F}}{16\pi} \right) \;-\; \frac{1}{4\pi}\;F^{\mu\alpha}\,F_{\alpha}^{\;\;\nu} \;-\;\left\langle
c\gamma^{\mu}\,D^{\nu}\right\rangle
\label{eq:T_Dirac}
\end{equation}
is derived by Noether method from the Dirac-Maxwell Lagrangian density ${\cal L} =  \langle c\gamma^{\mu}D_{\mu} - mc^{2}\rangle + F^{\mu\nu}F_{\nu\mu}/
16\pi$. While the electromagnetic terms in Eq.~\eq{T_Dirac} are symmetric, the Dirac term is not. Indeed, the antisymmetric part  $T_{\sf A}^{\mu\nu} \equiv \frac{1}{2} (T^{\mu\nu} - T^{\nu\mu})$ of Eq.~\eq{T_Dirac} is expressed as \cite{Hilgevoord}
\begin{equation}
T^{\mu\nu}_{\sf A} = -\,\frac{c}{2}\,\left\langle \gamma^{\mu}\,D^{\nu} \;-\frac{}{} \gamma^{\nu}\,D^{\mu}\right\rangle \equiv -\,\frac{1}{2}\,
\partial_{\alpha}\left\langle\Sigma^{\alpha[\mu\nu]}\right\rangle,
\label{eq:T_A_Dirac}
\end{equation}
where the third-rank tensor $\Sigma^{\alpha[\mu\nu]} \equiv -\,\Sigma^{\alpha[\nu\mu]}$:
\begin{equation}
\Sigma^{\alpha[\mu\nu]} \;\equiv\; \frac{c}{4}\,\left[ \left( S^{\alpha\mu}\,\gamma^{\nu} - S^{\alpha\nu}\,\gamma^{\mu}\right) + \left(
\gamma^{\mu}\,S^{\nu\alpha} - \gamma^{\nu}\,S^{\mu\alpha}\right)\right]
\label{eq:Sigma_Dirac} 
\end{equation}
is expressed in terms of the intrinsic spin angular momentum $S^{\alpha\beta} \equiv \hbar\,\sigma^{\alpha\beta}/2$. Since $(c\gamma^{i}\,D^{j} - 
c\gamma^{j}\,D^{i})$ is the operator-equivalent of the cross-product $\epsilon_{ijk}\,v^{i}\,p_{\rm kin}^{j} \equiv ({\bf v}\btimes{\bf p}_{\rm kin}
)_{k}$, the antisymmetry of the Dirac-Maxwell canonical energy-momentum tensor is, therefore, due to the decoupling of its velocity and kinetic momentum.

Belinfante \cite{Belinfante_1939,Belinfante_1940} recognized that the asymmetry of the canonical energy-momentum tensor $T^{\mu\nu}$ could be given a physical interpretation (see also Refs.~\cite{Rosenfeld,McLennan}), based on the fact that the transformation 
\begin{equation}
\ov{T}^{\mu\nu} \;\equiv\; T^{\mu\nu} \;+\; \partial_{\alpha}R^{[\alpha\mu]\nu}
\label{eq:ov_T_munu}
\end{equation}
leaves the energy-momentum conservation law invariant:
\begin{equation}
0 \;=\; \partial_{\mu}T^{\mu\nu} \;=\; \partial_{\mu}\ov{T}^{\mu\nu},
\label{eq:em_law}
\end{equation}
where the third-rank tensor $R^{[\mu\alpha]\nu} \equiv -\,R^{[\alpha\mu]\nu}$ satisfies $\partial^{2}_{\alpha\mu}R^{[\alpha\mu]\nu} \equiv 0$. The condition that the new energy-momentum tensor \eq{ov_T_munu} is symmetric yields the following expression for the antisymmetric part of the canonical energy-momentum tensor
\begin{equation}
T_{\sf A}^{\mu\nu} = -\,\frac{1}{2}\;\partial_{\alpha} \left( R^{[\alpha\mu]\nu} \;-\frac{}{} R^{[\alpha\nu]\mu} \right) \equiv -\,\frac{1}{2}\;
\partial_{\alpha}\,\Sigma^{\alpha[\mu\nu]},
\label{eq:T_A_def}
\end{equation}
and, hence, the antisymmetric part $T_{\sf A}^{\mu\nu}$ of the canonical energy-momentum tensor acts as the source for the antisymmetric second-rank tensor $\Sigma^{\alpha[\mu\nu]}$. 

We now show that the third-rank tensor $\Sigma^{\alpha[\mu\nu]}$ is connected to the spin angular-momentum tensor as in the Schroedinger 
{\it zitterbewegung} model [see Eq.~\eq{Sigma_Dirac}]. First, we introduce the third-rank orbital-angular-momentum tensor
\begin{equation}
L^{\beta[\mu\nu]} \;\equiv\; x^{\mu}\,T^{\beta\nu} \;-\; x^{\nu}\,T^{\beta\mu},
\label{eq:L_ang}
\end{equation}
and the third-rank spin-angular-momentum tensor
\begin{eqnarray}
S^{\beta[\mu\nu]} & \equiv & x^{\mu}\,\partial_{\alpha}R^{[\alpha\beta]\nu} \;-\; x^{\nu}\,\partial_{\alpha}R^{[\alpha\beta]\mu} \nonumber \\
 & = & \Sigma^{\beta[\mu\nu]} \;-\; \partial_{\alpha}Q^{[\alpha\beta][\mu\nu]},
\label{eq:S_ang}
\end{eqnarray}
where $Q^{[\alpha\beta][\mu\nu]} \equiv R^{[\alpha\beta]\mu}\,x^{\nu} - R^{[\alpha\beta]\nu}\,x^{\mu}$ satisfies $\partial^{2}_{\alpha\beta}
Q^{[\alpha\beta][\mu\nu]} \equiv 0$. Using Eq.~\eq{ov_T_munu}, the total angular-momentum tensor
\begin{equation}
J^{\beta[\mu\nu]} \;\equiv\; L^{\beta[\mu\nu]} \;+\; S^{\beta[\mu\nu]} \;=\; x^{\mu}\,\ov{T}^{\beta\nu} \;-\; x^{\nu}\,\ov{T}^{\beta\mu}
\label{eq:J_def}
\end{equation}
satisfies the angular-momentum conservation law
\begin{equation}
\partial_{\beta}J^{\beta[\mu\nu]} \;=\; \ov{T}^{\mu\nu} - \ov{T}^{\nu\mu} \;\equiv\; 0,
\label{eq:ang_mom_cons}
\end{equation}
which follows from the symmetry of the new energy-momentum tensor \eq{ov_T_munu}. Lastly, the equation $\partial_{\beta}L^{\beta[\mu\nu]}$ for the orbital-angular-momentum tensor \eq{L_ang} is
\begin{equation}
\partial_{\beta}L^{\beta[\mu\nu]} = T^{\mu\nu} - T^{\nu\mu} \equiv -\,\partial_{\beta}S^{\beta[\mu\nu]} = -\,\partial_{\beta}\Sigma^{\beta[\mu\nu]}.
\label{eq:L_motion}
\end{equation}
Hence, we see that the asymmetry of the canonical energy-momentum tensor acts as the source of intrinsic (spin) angular momentum.

We note that the antisymmetric energy-momentum tensor
\begin{equation}
T_{\sf A}^{\mu\nu} \;=\; -\,\frac{1}{2}\;\varepsilon^{\mu\nu\alpha\beta}\;\partial_{\alpha}\sigma_{\beta} \;\equiv\; \frac{1}{4}\;
\varepsilon^{\mu\nu\alpha\beta}\;\tau_{\alpha\beta}
\label{eq:TA_tau}
\end{equation}
can be used to define the antisymmetric torque tensor $\tau_{\alpha\beta} \equiv -\,(\partial_{\alpha}\sigma_{\beta} - \partial_{\beta}
\sigma_{\alpha})$, with the spatial components $T_{\sf A}^{ij} \equiv \frac{1}{2}\,\varepsilon^{0ijk}\,\tau_{0k}$ expressed in terms of the intrinsic torque tensor $\tau_{0\mu} \equiv (0, \vb{\tau})$:
\begin{equation}
\vb{\tau} \;\equiv\; {\bf v}\btimes{\bf p} \;+\; {\bf E}\btimes{\bf P} \;+\; {\bf B}\btimes{\bf M}.
\label{eq:torque_vec}
\end{equation}
We therefore see that all three features of the {\it zitterbewegung} model are combined in Eq.~\eq{torque_vec} to act as the source of an intrinsic (spin) angular momentum.

After a brief introduction to the Schroedinger {\it zitterbewegung} model and a general discussion of angular-momentum conservation within a Lagrangian perspective, we now discuss the case of the Lagrangian formulation of reduced Vlasov-Maxwell theory. The Lagrangian formulation of the guiding-center and oscillation-center Vlasov-Maxwell equations were developed over 30 years ago \cite{Dewar,Cary_K,PM_85,PLS,Boghosian,Ye_K}. In each case, their respective energy-momentum conservation laws, derived by Noether method, exhibited an asymmetric canonical energy-momentum tensor. Although it was sometimes pointed out that this asymmetry was only apparent \cite{PLS}, based on the knowledge that the conservation of angular momentum required a symmetric physical energy-momentum tensor, Dewar \cite{Dewar} pointed out that the asymmetry of the canonical energy-momentum tensor could be expressed in terms of an intrinsic spin angular-momentum tensor. 

The process of dynamical reduction in single-particle plasma dynamics and plasma kinetic theory \cite{Brizard_Vlasovia} is associated with the extended near-identity canonical phase-space transformation $\cal{T}_{\epsilon}: \wh{z}^{a} = z^{a} + \epsilon\;\{ S_{1}, z^{a}\} + \epsilon^{2}\,(\{ 
S_{2}, z^{a}\} + \frac{1}{2}\,\{ S_{1}, \{ S_{1}, z^{a}\}\}) + \cdots$, and its inverse $\cal{T}_{\epsilon}^{-1}: z^{a} = \wh{z}^{a} - \epsilon\;\{ 
S_{1}, \wh{z}^{a}\} - \epsilon^{2}\,(\{ S_{2}, \wh{z}^{a}\} - \frac{1}{2}\,\{ S_{1}, \{ S_{1}, \wh{z}^{a}\}\}) + \cdots$, generated by the scalar fields
$(S_{1}, S_{2},\cdots)$. This near-identity transformation introduces the reduced-displacement vector 
\begin{equation}
\vb{\rho}_{\epsilon} \;\equiv\; {\sf T}_{\epsilon}^{-1}{\bf x} \;-\; \wh{{\bf x}} \;=\; -\,\epsilon\;G_{1}^{{\bf x}} + \cdots, 
\label{eq:rho_epsilon}
\end{equation}
defined as the difference between the push-forward ${\sf T}_{\epsilon}^{-1}{\bf x}$ of the particle position ${\bf x}$ and the reduced position 
$\wh{{\bf x}}$. Hence, through the reduced displacement \eq{rho_epsilon}, the dynamical reduction yields the reduced electric-dipole moment 
\cite{Brizard_Vlasovia}
\begin{equation}
\vb{\pi}_{\epsilon} \;\equiv\; e\;\vb{\rho}_{\epsilon}
\label{eq:pi_def}
\end{equation}
and the intrinsic magnetic-dipole moment
\begin{equation}
\vb{\mu}_{\epsilon} \;\equiv\; \frac{e}{2c}\;\left(\vb{\rho}_{\epsilon}\btimes \frac{d_{\epsilon}\vb{\rho}_{\epsilon}}{dt}\right),
\label{eq:mu_def}
\end{equation}
which is identical in form to the {\it zitterbewegung} expression \eq{mu_electron} for the electron's magnetic-dipole moment.

The extended reduced Vlasov equation for the extended reduced Vlasov distribution $\wh{{\cal F}}$ is expressed as
\begin{equation}
0 \;=\; \frac{d_{\epsilon}\wh{{\cal F}}}{dt} \;\equiv\; \frac{d_{\epsilon}\wh{z}^{a}}{dt}\;\pd{\wh{{\cal F}}}{\wh{z}^{a}},
\label{eq:redextVlasov_def}
\end{equation}
where
\begin{equation}
\frac{d_{\epsilon}\wh{x}^{\mu}}{dt} \;=\; \pd{\wh{\cal H}}{\wh{p}_{\mu}} \;\;{\rm and}\;\; \frac{d_{\epsilon}\wh{p}_{\mu}}{dt} \;=\; -\;
\pd{\wh{\cal H}}{\wh{x}^{\mu}}
\label{eq:can_z_dot}
\end{equation}
denote the canonical Hamilton equations in extended phase space and the reduced Vlasov distribution is defined as $\wh{{\cal F}}(\wh{{\sf z}}) \equiv 
c\,\delta[\wh{w} - \wh{H}(\wh{{\bf x}}, \wh{{\bf p}}, t)]\, \wh{F}(\wh{{\bf x}}, \wh{{\bf p}}, t)$, with the reduced extended Hamiltonian satisfying the physical constraint $\wh{{\cal H}} = \wh{H} - \wh{w} \equiv 0$, where the reduced Hamiltonian
\begin{equation}
\wh{H} \equiv H - \epsilon\,\frac{dS_{1}}{dt} - \epsilon^{2} \left( \frac{dS_{2}}{dt} - \frac{1}{2} \left\{ S_{1},\; \frac{dS_{1}}{dt}\right\}
\right) + \cdots
\label{eq:ovH_def}
\end{equation}
is expressed in a form that is the classical equivalent of the Dirac Hamiltonian derived by the Foldy-Wouthuysen (FW) transformation \cite{RQM}. 

The dynamical reduction associated with the phase-space transformation $\cal{T}_{\epsilon}$ introduces polarization and magnetization effects into the Maxwell equations, which are transformed into the macroscopic (reduced) Maxwell equations \cite{Ye_K,Brizard_Vlasovia}
\begin{equation}
\nabla\bdot{\bf D} \;=\; 4\pi\,\wh{\varrho} \;\;{\rm and}\;\;
\nabla\btimes{\bf H} \;-\; \frac{1}{c}\,\pd{{\bf D}}{t} \;=\; \frac{4\pi}{c}\,\wh{{\bf J}},
\label{eq:DH_eq}
\end{equation} 
where the microscopic electric and magnetic fields ${\bf E}$ and ${\bf B}$ are replaced by the macroscopic fields ${\bf D} \equiv {\bf E} + 4\pi\,
{\bf P}_{\epsilon}$ and ${\bf H} \equiv {\bf B} - 4\pi\,{\bf M}_{\epsilon}$, where ${\bf P}_{\epsilon}$ and ${\bf M}_{\epsilon}$ are the reduced polarization and magnetization. We note that the dynamical reduction associated with the phase-space transformation $\cal{T}_{\epsilon}$ has introduced the following expressions for the charge and current densities:
\begin{equation}
\left. \begin{array}{rcl}
\varrho & \equiv & \wh{\varrho} \;-\; \nabla\bdot{\bf P}_{\epsilon} \\
{\bf J} & \equiv & \wh{{\bf J}} \;+\; \partial{\bf P}_{\epsilon}/\partial t \;+\; c\,\nabla\btimes{\bf M}_{\epsilon}
\end{array} \right\}, 
\label{eq:rhoJ_Rpolmag}
\end{equation}
which are of course similar to the Gordon decomposition \eq{rho_J_elec} observed in the Dirac model.

Lastly, by using the reduced electric-dipole and magnetic-dipole moments \eq{pi_def}-\eq{mu_def}, we construct explicit expressions for the reduced polarization
\begin{equation}
{\bf P}_{\epsilon} \;\equiv\; \sum\;\int\;\vb{\pi}_{\epsilon}\;\wh{F}\,d^{3}\wh{p},
\label{eq:red_pol}
\end{equation}
and the reduced magnetization
\begin{equation}
{\bf M}_{\epsilon} \;\equiv\; \sum\;\int\;\left( \vb{\mu}_{\epsilon} \;+\; \frac{\vb{\pi}_{\epsilon}}{c}\btimes\frac{d_{\epsilon}\wh{\bf x}}{dt}
\right)\;\wh{F}\,d^{3}\wh{p},
\label{eq:red_mag}
\end{equation}
which combines the intrinsic magnetic-dipole contribution \eq{mu_def} and the moving electric-dipole contribution.

We now show that the reduced Vlasov-Maxwell equations \eqref{eq:redextVlasov_def} and \eqref{eq:DH_eq} can be derived from the reduced variational principle $\int d^{4}x\;\delta\wh{\cal{L}} = 0$, where the reduced Lagrangian density is
\begin{eqnarray}
\wh{\cal{L}}(x) & \equiv & -\; \sum\;\int d^{4}\wh{p}\;\wh{{\cal F}}(x,\wh{p})\;\wh{{\cal H}}(x,\wh{p}; A, {\sf F}) \nonumber \\
 &  &+\; \frac{1}{16\pi}\,{\sf F}(x):{\sf F}(x).
\label{eq:redLag_def}
\end{eqnarray}
Note that, as a result of the dynamical reduction of the Vlasov equation, the reduced Hamiltonian \eq{ovH_def} is expressed in terms of canonical energy-momentum coordinates as
\begin{eqnarray}
\wh{H}(\wh{\bf x}, \wh{\bf p},t; A, {\sf F}) & \equiv & \frac{1}{2m}\,|\wh{{\bf p}} - (e/c)\,{\bf A}|^{2} \;+\; e\,\Phi \nonumber \\
 &  &+\; \Psi_{\epsilon}\left(\wh{{\bf p}} - \frac{e}{c}\,{\bf A}; {\sf F}\right),
\label{eq:ovH_def}
\end{eqnarray}
where the reduced {\it ponderomotive} potential $\Psi_{\epsilon}$ depends explicitly on the field tensor ${\sf F}_{\mu\nu}$, which once again is the classical equivalent of the Dirac Hamiltonian after the FW transformation. From these field dependences, we define the reduced four-current density
\begin{equation}
\wh{J}^{\mu} \;=\; (c\wh{\varrho}, \wh{{\bf J}}) \;\equiv\; \sum\;e\;\int d^{4}\wh{p}\;\wh{{\cal F}}\;\frac{d_{\epsilon}\wh{x}^{\mu}}{dt},
\label{eq:redJ_var}
\end{equation}
where $d_{\epsilon}\wh{x}^{0}/dt = c$ and
\begin{equation}
\frac{d_{\epsilon}\wh{\bf x}}{dt} \;\equiv\; \pd{\wh{{\cal H}}}{\wh{\bf p}} \;=\; \frac{1}{m}\,(\wh{{\bf p}} - \frac{e}{c}\,{\bf A}) + 
\pd{\Psi_{\epsilon}}{\wh{{\bf p}}}.
\label{eq:dx_epsilon}
\end{equation}
The reduced antisymmetric polarization-magnetization tensor \cite{Boghosian}, on the other hand, yields the reduced polarization ${\sf K}_{\epsilon}^{0i} = P_{\epsilon}^{i}$ and the reduced magnetization ${\sf K}_{\epsilon}^{ij} = \epsilon^{ijk}\,M_{\epsilon\;k}$, where 
\begin{equation}
\left( {\bf P}_{\epsilon},\; {\bf M}_{\epsilon} \right) \;\equiv\; -\;\sum\;\int d^{4}\wh{p}\;\wh{{\cal F}}\;
\left( \pd{\Psi_{\epsilon}}{{\bf E}},\; \pd{\Psi_{\epsilon}}{{\bf B}} \right). 
\label{eq:PM_var}
\end{equation}
Hence, polarization and magnetization are explicitly evaluated in terms of the ponderomotive Hamiltonian $\Psi_{\epsilon}$.

By applying the Nother method, we obtain the reduced canonical energy-momentum tensor
\begin{eqnarray}
{\sf T}^{\mu\nu} & \equiv & \frac{1}{4\pi} \left[\; \frac{g^{\mu\nu}}{4}\; {\sf F}:{\sf F} \;-\; \left( {\sf F}^{\mu\sigma} \;+\frac{}{} 4\pi\,
{\sf K}_{\epsilon}^{\mu\sigma}\right)\;{\sf F}_{\sigma}^{\;\;\nu} \;\right] \nonumber \\
 &  &+\; \sum\;\int d^{4}\wh{p}\; \pd{\wh{{\cal H}}}{\wh{p}_{\mu}}\;\left(\wh{p}^{\nu} - \frac{e}{c}\,A^{\nu} \right)\;\wh{{\cal F}}, 
\label{eq:Tmunu_def}
\end{eqnarray}
which naturally includes the polarization and magnetization \eqref{eq:PM_var} and is manifestly asymmetric. The antisymmetric part of the canonical energy-momentum tensor $T_{\sf A}^{ij} \equiv \frac{1}{2}\,\epsilon^{ijk}\,\tau_{k}$ is expressed in terms of the reduced (intrinsic) {\it ponderomotive} torque
\begin{eqnarray}
\vb{\tau} & \equiv & \sum\,\int\,\left[ \pd{\Psi_{\epsilon}}{\bf E}\btimes{\bf E} \;+\; \pd{\Psi_{\epsilon}}{\bf B}\btimes{\bf B} \right. \nonumber \\
 &  &\left.\hspace*{0.5in}+\; \pd{\Psi_{\epsilon}}{\wh{\bf p}}\btimes\left(\wh{\bf p} - \frac{e}{c}\,{\bf A} \right) \right]\;\wh{F}\, d^{3}\wh{p}.
\label{eq:torque_pond}
\end{eqnarray}
The three {\it zitterbewegung} features, which appear explicitly in the expression for the reduced ponderomotive torque, are derived from the reduced ponderomotive potential $\Psi_{\epsilon}$ associated with the process of dynamical reduction.

As an application of the ponderomotive torque \eq{torque_pond}, we consider the gyrocenter ponderomotive Hamiltonian \cite{Brizard_energy}
\begin{equation}
\Psi_{\epsilon} \;=\; -\;\frac{mc^{2}}{2\,B_{0}^{2}}\; \left|\,{\bf E}_{\bot} \;+\; \frac{\wh{p}_{\|}\bhat_{0}}{mc}\btimes{\bf B}_{\bot}\,\right|^{2}
\;+\; \mu\;\frac{|{\bf B}_{\bot}|^{2}}{2B_{0}},
\label{eq:Psi_gy}
\end{equation}
derived in the zero-Larmor-radius (ZLR) limit from the second-order gyrocenter Hamiltonian \cite{Brizard_Hahm}, where a strongly magnetized background plasma is perturbed by the low-frequency electromagnetic fields ${\bf E}_{\bot} \equiv -\nabla_{\bot}\Phi$ and ${\bf B}_{\bot} \equiv \nabla_{\bot}A_{\|}\btimes\bhat_{0}$, which are perpendicular to the background magnetic field ${\bf B}_{0} = B_{0}\,\bhat_{0}$. From the gyrocenter ponderomotive Hamiltonian \eq{Psi_gy}, we obtain the reduced gyrocenter polarization and magnetization
\begin{eqnarray}
\pd{\Psi_{\epsilon}}{{\bf E}_{\bot}} & = & -\;\frac{mc^{2}}{B_{0}^{2}} \left( {\bf E}_{\bot} \;+\; \frac{\wh{p}_{\|}\bhat_{0}}{mc}\btimes{\bf B}_{\bot}
\right) \;\equiv\; -\,\vb{\pi}_{\epsilon}, \label{eq:Psi_gy_E} \\
\pd{\Psi_{\epsilon}}{{\bf B}_{\bot}} & \equiv & -\,\vb{\mu}_{\epsilon} \;-\; \vb{\pi}_{\epsilon}\btimes \frac{\wh{p}_{\|}\bhat_{0}}{mc}, 
\label{eq:Psi_gy_B}
\end{eqnarray}
where $\vb{\mu}_{\epsilon} \equiv -\,\mu\,{\bf B}_{\bot}/B_{0}$ denotes the reduced intrinsic gyrocenter magnetization. These definitions lead to the identity $\partial\Psi_{\epsilon}/\partial{\bf E}_{\bot}\btimes{\bf E}_{\bot} = {\bf E}_{\bot}\btimes\vb{\pi}_{\epsilon} \equiv -\;\partial
\Psi_{\epsilon}/\partial{\bf B}_{\bot}\btimes{\bf B}_{\bot}$, which means that the electromagnetic contribution to the gyrokinetic intrinsic torque vanishes in the ZLR limit. The kinetic part of the gyrokinetic intrinsic torque, on the other hand, involves the ponderomotive velocity
\begin{eqnarray}
\pd{\Psi_{\epsilon}}{\wh{\bf p}} & = & -\,\bhat_{0} \left[ \frac{{\bf B}_{\bot}}{B_{0}}\bdot \left( {\bf E}_{\bot}\btimes\frac{c\bhat_{0}}{B_{0}} \;+\; \frac{\wh{p}_{\|}}{m}\;\frac{{\bf B}_{\bot}}{B_{0}} \right) \right] \nonumber \\
 & \equiv & -\,\bhat_{0}\;\left|\pd{\Psi_{\epsilon}}{\wh{\bf p}}\right|,
\label{eq:Psi_p}
\end{eqnarray}
which vanishes only in the electrostatic limit $({\bf B}_{\bot} \equiv 0)$. Lastly, the intrinsic gyrokinetic torque 
\begin{equation}
\vb{\tau}_{\rm gy} \;=\; \sum\,\int\,\left( \left|\pd{\Psi_{\epsilon}}{\wh{\bf p}}\right|\;\wh{\bf p}\btimes\bhat_{0} \right)\;\wh{F}\, d^{3}\wh{p}
\label{eq:tau_gy}
\end{equation}
involves the magnitude of the ponderomotive velocity \eq{Psi_p} and the perpendicular component of the gyrocenter kinetic momentum $[\wh{\bf p} - (e/c)\,
A_{\|}\,\bhat_{0}]\btimes\bhat_{0} = \wh{\bf p}\btimes\bhat_{0}$.

In summary, we have shown how classical {\it zitterbewegung} effects manifest themselves in reduced Vlasov-Maxwell theory through the reduced ponderomotive Hamiltonian. Here, the ponderomotive torque \eq{torque_pond} acts as the source for the intrinsic spin angular momentum. Ongoing work includes the investigation of classical {\it zitterbewegung} effects in gyrofluid and gyrokinetic models.

\end{document}